\documentclass[fleqn,usenatbib,usedcolumn]{mnras}

\usepackage{newtxtext,newtxmath}

\usepackage[T1]{fontenc}
\usepackage{ae,aecompl}

\usepackage{graphicx}	

\newcommand{\ltsima}{$\; \buildrel < \over \sim \;$}
\newcommand{\simlt}{\lower.5ex\hbox{\ltsima}}
\newcommand{\gtsima}{$\; \buildrel > \over \sim \;$}
\newcommand{\simgt}{\lower.5ex\hbox{\gtsima}}

\title[Blue Straggler - Metallicity Connection]{The Rich Lack Close Neighbours: The Dependence of Blue-Straggler Fraction on Metallicity}

 \author[Wyse et al.]{Rosemary F. G. Wyse$^{1,2}$\thanks{E-mail: wyse@jhu.edu (RFGW)}, Maxwell Moe$^{3}$  and  Kaitlin M.~ Kratter$^{3}$  \\
   $^{1}$Department of Physics \& Astronomy, Johns Hopkins University, 3400 N.~Charles Street, Baltimore, MD 21218, USA \\
   $^{2}$Kavli Institute for Theoretical Physics, University of California, Santa Barbara, CA 93106, USA \\
$^{3}$Steward Observatory, University of Arizona, 933 N.~Cherry Ave., Tucson, AZ 8572,  USA 
 }

\date{Accepted XXX. Received YYY; in original form ZZZ}

\pubyear{2020}

\begin{document}
\label{firstpage}
\pagerange{\pageref{firstpage}--\pageref{lastpage}}
\maketitle

\begin{abstract}
  Blue straggler stars (BSS) have been identified in star clusters and in field populations in our own Milky Way galaxy and in its satellite galaxies.  They manifest as stars bluer and more luminous than the dominant old population, and usually have a spatial distribution that follows the old population. Their progenitors are likely to have been close binaries. We investigate trends of the BSS population in dwarf spheroidal galaxies (dSph) and in the bulge of the Milky Way and find an anti-correlation between the relative frequency of BSS and the metallicity of the parent population. The rate of occurrence of BSS in the metal-poor dwarf galaxies is approximately twice that found in the solar-metallicity bulge population.  This trend of decreasing relative population of BSS  with increasing metallicity  mirrors that found for the close-binary fraction in the field population of the Milky Way.  We argue that the dominant mode of BSS formation in low-density environments is likely to be mass transfer in close-binary systems. It then follows that the similarity between the trends for BSS in the dSph and field stars in our Galaxy supports the proposal that the small-scale fragmentation during star formation is driven by the same dominant  physical process, despite the diversity in environments,  plausibly gravitational instability of proto-stellar discs.  
\end{abstract}





\section{Introduction}

\label{sec:Introduction}
Blue straggler stars (BSS) were first identified as stars bluer and
brighter than the dominant (old) main-sequence turn-off in the
colour-magnitude diagram (CMD) of the globular cluster M3
\citep{sandage1953}. It is now well-established
\citep[e.g.][]{Piotto2004} that globular clusters contain rich
populations of blue stragglers. In the absence of supporting evidence
for a genuinely younger (sub)population, BSS are thought be formed
through a rejuvenation of a main-sequence member of the old
population, via two favoured mechanisms: either mass transfer in a
close binary \citep{McCrea1964} or dynamically induced stellar
collisions and mergers \citep{hills1976}. Both mechanisms are expected
to operate in globular clusters \citep[e.g.][]{hut1983,bailyn1995}
with the relative importance of each being a function of the cluster's
structure and dynamical state. The relative concentration of the BSS
population has been proposed as a measure of the dynamical age of the
globular cluster \citep[e.g.][]{ferraro2018}. Open clusters can also
host BSS \citep[e.g.][]{milone1994,AL2007}. Indeed luminous old open
clusters, which presumably survived many orbital times due to being sufficiently dense, 
are dynamically evolved enough  that both formation channels of
BSS could have operated
\citep[e.g.][]{leonard1996,hurley2005,tian2006,chen2009,geller2016}. At least in the old open cluster NGC~188 (age $\sim 7$~Gyr), the majority of the BSS are in binary systems - as high as 80\% of the BSS are spectroscopic binaries in the survey of \citet{mathieu2009}, with the companion likely to be a white dwarf. This persistence of binaries implies that internal dynamical processes have not induced global stellar mergers/collisions, modulo that many of the close binaries could have evolved in a triple system \citep{triples2009}.    BSS populations in
clusters depend on the properties of (primordial) binary systems,
binary star evolution, stellar evolution and dynamical processes in
dense stellar environments, with ongoing debate as to the dominant
mechanism \citep[e.g.][]{sills2018}. 

Blue stragglers have also been identified in the field populations of
the Milky Way, in particular among Blue Metal-Poor (BMP) stars in the
stellar halo which show a high binary fraction 
\citep[$\sim 60$\%,][]{preston2000}). More recently the BSS population in the field
stellar halo and thick disc was estimated from analyses of the SDSS
photometric and spectroscopic data for A-type stars \citep{santucci2015}. These stars consist of both giants (blue
horizontal-branch (BHB) stars) and main-sequence dwarfs (candidate BSS
plus foreground intermediate-age disc stars).   BSS
in the field are expected to form predominantly from mass transfer in
binaries, providing insight into the properties of binary systems at
early times. The different technique used to identify the BSS in the
field old stellar population of the Milky way, namely by spectroscopic
estimation of stellar gravity (BSS have higher gravity than do
horizontal branch stars) rather than by location in the CMD, makes
comparisons with the relative frequency of BSS to BHB measured in
clusters is difficult. Further complications in quantifying the relative frequency BSS/BHB in the halo and thick disc arise from
the uncertain absolute magnitudes of the BSS in the field and the
different volume elements probed by the dwarf and giant A-stars (as
discussed in \citealt{santucci2015}.

The BSS candidates in the field stellar halo identified by \citet{preston2000} are predominantly spectroscopic binaries, with orbital periods of less than 3000~days (see their Table~5; as they discuss, a few longer period systems could exist). A recent analysis of the binarity of solar-mass stars (primary with mass in the range of 0.6~M$_\odot$ to 1.5~M$_\odot$) in the Milky
Way Galaxy, paying particular attention to completeness corrections, revealed a significant anti-correlation between the 
fraction of stars in close binaries  (periods less than $10^4$ days, semi-major axis less than 10~AU) and their metallicity
\citep[]{Moe2018}. This sample of Galactic binary systems contains stars in a range of evolutionary stages and primarily probes the
disc and halo, spanning a metallicity range from $ \sim -3$~dex to
$\sim +0.5$~dex.  These authors found that the fraction of solar-type close binaries (separations less than 10~AU) equals 53\%$\pm$12\% in a parent stellar population of mean metallicity <[Fe/H]> $ = -3.0$,  40\%$\pm$6\% at <[Fe/H]> $ = -1.0$,  24\%$\pm$4\% at <[Fe/H]> $ = -0.2$ and 10\%$\pm$3\% at  <[Fe/H]> $ = +0.5$. This trend is consistent with that found by \citet{badenes2018} for low-mass stars, based on radial-velocity variation within the public data from the APOGEE survey.

The anti-correlation between close-binary fraction and metallicity may arise from enhanced  small-scale fragmentation of 
proto-stellar discs at lower metallicities, particularly for
low-mass ($\sim 1~M_\odot$) protostars \citep[]{Moe2018}. The analysis of the binary-star population within  {\it Gaia\/} DR2 by \citet{elbadry2019} indeed revealed an anti-correlation between binary fraction and metallicity for close binaries, while there is little or no dependence on metallicity for the wide-binary fraction.  \citet{elbadry2019} estimate the scale at which the anti-correlation sets in is for semi-major axes below  around 100-200~AU. They find consistency between  the amplitude of the anti-correlation they find  for the smallest separations in their sample (between 50~AU and 100~AU) and that found by \citet{Moe2018} (see  Fig.~2 in \citealt{elbadry2019}).

As noted above, stars in close binaries can undergo mass transfer and may be progenitors of BSS. 
One might then expect to find an
anti-correlation between the blue-straggler fraction and metallicity
of the parent population, provided mass transfer in binary systems is the dominant mode of
blue-straggler formation.  The  present
frequency of occurrence of BSS in  clusters is not expected to correlate exactly with the initial close-binary
fraction, due to internal dynamical effects \citep[e.g.][]{hut1983}. 
Lower-density stellar systems that contain BSS should provide a more robust examination of possible trends in the relative frequency of a primordial close-binary population, as the main channel for BSS in these environments should be mass-transfer in primordial binary systems. BSS are most easily
determined for parent populations with a narrow age range. Several of the dwarf spheroidal (dSph) satellite galaxies of the 
Milky Way have predominantly old stellar populations and indeed analyses of their CMDs  reveal stars bluer and more luminous than the dominant old main-sequence turnoff  \citep[e.g.][]{Momany2007,santana2013}. Under the hypothesis that these are BSS that arise from mass transfer in close binaries, their relative frequency as a function of the host metallicity can
provide an independent - and extragalactic -  test of the universality
of a metallicity
dependence of the close binary fraction, and of the underlying physical mechanism. In this paper we analyse existing data for dSph, taken from the
literature, to determine how the relative frequency of BSS  depends on the metallicity of the host galaxy. We extend the metallicity range by inclusion of the Galactic bulge, where BSS are also identified through their location in the CMD.

\section{The Blue Straggler Populations in Old Stellar Systems} \label{sec:data}
\subsection{Classical Dwarf Spheroidal Galaxies}
\label{subsec:dSph}

Dwarf spheroidal galaxies (dSph) are characterised, in part, by their
low stellar surface densities and old stellar ages \citep[e.g.~review
  of][]{JGRW94}. This notwithstanding, analyses of their
colour-magnitude diagrams (CMDs) have revealed that many contain
populations of BSS. The `classical' dwarf spheroidal galaxies have a central surface brightness typically around 26~magnitudes/square arcsecond in the V-band, or $~1.5$~L$_\odot$/pc$^2$
\citep{IH95}. Such a low stellar density also would favour the possibility that the
blue-straggler population reflects a primordial population of close binaries \citep{McCrea1964}, rather than the collisional paths more likely to occur within the central regions of globular clusters \citep[e.g.~see discussion in][]{Mapelli2006}.

The identification
and quantification of these `stragglers' is cleaner for parent
`simple' stellar populations, i.e.~narrow ranges of ages and
metallicities (in addition, at low metallicities, the location in the
CMD of an old turnoff does not vary much with metallicity). Happily,
many of the classical dwarf spheroidal satellite galaxies of the Milky Way do
satisfy these conditions, although the possibility of the BSSs  being {younger stars, with ages down to $\simlt 1$~Gyr, still requires careful
consideration,} as briefly discussed below \citep[see also the reviews
  of][]{sills1,sills2}.

The Ursa Minor dSph is a well-studied, relatively nearby, `classical' dwarf spheroidal (heliocentric
distance of $\sim 75$~kpc, \citep{Carrera2002}).  A possible blue straggler
population in the Ursa Minor dSph was identified,  by
\citet{OA85}, in ground-based imaging that
reached just below the dominant old main-sequence turn-off.  Subsequent deep (small-area) imaging with the Hubble Space
Telescope provided a cleaner CMD and again identified a
blue straggler population \citep{RFGW2002}.  The probable  nature of this population as BSS, as opposed to genuinely younger stars, was established from wide-area ground-based imaging (out to the tidal limit) by
\citet{Carrera2002}, using arguments based largely on their spatial distribution, which follows that of the old population \citep[see also][]{Mapelli2007}.

The
Draco dSph similarly consists of old stars with a narrow age range
\citep{Dolphin02,Mapelli2007}.  The Sculptor dSph also has a dominant old population
\citep{Dolphin02}, however with a centrally concentrated younger and more
metal-rich component \citep[ages $\sim 8$~Gyr][]{deboer-scl}. A similar population
gradient is seen in the Sextans dSph \citep{okamotosex,cic2018}. Blue straggler
populations are nonetheless identifiable in all of these galaxies, as stars significantly bluer and brighter than the dominant old turn-offs, and with  radial profiles consistent with those of the older populations \citep[e.g.][]{Momany2007, Mapelli2009, okamotosex, cic2018}. {The Sagittarius dwarf is discussed in section~\ref{sec:Sgr} below and the Carina dSph in section~\ref{sec:carina}.}

\subsection{Ultra-Faint Dwarf Spheroidal Galaxies}
\label{subsec:UFD}

The advent of wide-area imaging surveys
with uniform, well-calibrated photometry led to the discovery of
`ultra-faint' dSph \citep{FoS, BooI} that have total luminosities equal to those of faint star clusters and  colour-magnitude
diagrams consistent with old, metal-poor populations. Subsequent deep
imaging, combined with spectroscopic metallicity determinations,
confirmed that these low-luminosity, extremely low stellar surface-density
systems indeed host ancient, low metallicity stellar populations \citep[e.g.][]{Norris2010, Brown2012, okamoto2012}. Those with sufficiently deep and wide data  were shown also to host  blue stragglers, again identified as stars bluer and more luminous than the dominant old main sequence turnoff, with a radial profile following that of the old stars \citep{Momany2007, okamoto2012, santana2013}.  
{The Willman 1 object  is discussed in section~\ref{sec:Wil1} below.}

\section{Specific Frequency of Blue Straggler Stars}
\label{sec:FBSS}

\subsection{Estimates Determined From Colour-Magnitude Diagrams}

The relative fraction of BSS has been estimated using a range of
definitions.  \citet{Carrera2002} quantified the blue-straggler
population relative to the number of horizontal-branch stars.
\citet{Mapelli2007} estimated the fraction of BSS relative both to
horizontal-branch stars and to red-giant stars. These latter authors noted the difficulty of
direct comparisons with estimates by others, which are often based on
photometric datasets in different bandpasses, adopting different colour and magnitude cuts to isolate BSS 
within the CMDs. This non-uniformity from author-to-author
means that trends of the relative population of blue
stragglers with properties of the host galaxy, such as those discussed
below, are best studied within any one analysis. It should be noted,
however, that there is good agreement across different analyses that
the  fraction of BSS, however defined, relative to the dominant old population shows a flat radial profile, indicating
that the blue stragglers are also an old stellar population.

There are two published studies of BSS that each provides a uniform
analysis of the relative populations of BSS to a tracer of the
dominant old population, in a representative sample of the satellite
galaxies of the Milky Way, namely that of \citet{santana2013}, who
derived estimates of the specific frequencies of BSS relative to the
red giant population, and that of \citet{Momany2007}, who derived
estimates relative to the horizontal branch (summing both red and blue
horizontal-branch stars).  Each of these two studies found BSS in all
the galaxies in their sample. The reported fractions of BSS from these
two analyses are shown in Table~\ref{tab:santana} and
Table~\ref{tab:momany}, respectively, together with   the mean stellar iron-peak abundance (plus error on the mean\footnote{The dispersion
around this mean value is generally significantly higher than
the error in the mean (and may increase systematically as the mean
metallicity decreases;  \citealt{Norris2010}.)} 
and absolute V-band magnitude (plus error) of the host galaxy, from
the compilation of \citet{AMcC2012}. The iron-peak 
abundances for the stars in these galaxies are derived from either moderate-resolution spectra \citep{Norris2010,kirby2011} or high-resolution spectra \citep[Segue~II and Sagittarius,][respectively]{belokurov2009,chou2007}.

The derived values of the fraction of BSS to RGB (F$_{RGB}^{BSS}$) are comparable to the quoted uncertainties  for several of the galaxies in the sample of \citet{santana2013}. We recalculated the values and uncertainties in terms of log(F$_{RGB}^{BSS}$) according to Poisson statistics, given the numbers of BSS and RGB stars given in \citet[their Table~1 and equation (1)]{santana2013}. These recalculated values are also given in Table~\ref{tab:santana} {(columns~3 and 4)} and are the quantities used below.  {\citet{Momany2007} give their derived values and errors already in logarithmic form and are used here as reported.} 

{The photometric data analysed by \citet{santana2013} were obtained as part of a new survey and consist of images in the $g-$ and $r$-band, taken with MegaCam on the Canada-France-Hawaii telescope. The data uniformly reach to at least one magnitude below the dominant main-sequence turnoff and cover out to, or beyond, two times the half-light radius of each system. }

{The photometric data analysed by \citet{Momany2007} were mostly obtained by others and made available to those authors, with the exception of Leo~II and the Sagittarius dSph, for which \citet{Momany2007} reduced and analysed raw archival data. The data they analysed are primarily in BVI optical broadband filters (see their Figure~1) and  in general extend to at least the nominal half-light radius of each galaxy. The Sagittarius dwarf is the only system for which the data are restricted to a small fraction of the galaxy (one square degree, excluding $14^\prime \times 14^\prime$ around the globular cluster M54, thus covering $\sim 6$\% of the main body, \citealp{Momany2007}). The colour-magnitude diagram for this galaxy also has significant contamination from Galactic stars (see Fig.~1 of \citealt{Momany2007}), resulting in larger uncertainties. }

\citet{Momany2007} found an anti-correlation between the relative
frequency of BSS and the luminosity of the host galaxy (their
Fig.~2). The sense of the correlation, decreasing frequency of BSS with increasing luminosity of the host system,  follows that established earlier
for globular clusters \citep{Piotto2004}, but the slope is less steep.
An apparently disparate  result was reported by \citet{santana2013}, who
found an essentially flat slope for the trend of the specific
frequency of BSS as a function of host galaxy luminosity, when using the RGB to normalise the number of BSS (their Fig.~4). The trend {with luminosity in \citet{Momany2007} is anchored at the bright end by the Sagittarius dSph, which is not included in the sample of \citet{santana2013}. }

\subsection{Individual  Systems with Unusual Characteristics}

  \subsubsection{Sagittarius dSph}
  \label{sec:Sgr}
  
  {We retained the Sagittarius dwarf spheroidal (Sgr) in the sample of \citet{Momany2007}. While this satellite galaxy is currently merging with the Milky Way and the outer parts have been stretched into tidal streams across the sky, the main body of the system shows no kinematic disturbance \citep{ibata97, frinch2012}. Further, although  some star formation occurred over many Gyr, there is a clearly dominant old--intermediate-age population. The Sgr dSph is the most luminous and most  metal-rich dwarf satellite galaxy in the sample. }
  
{The first deep photometric dataset for Sgr \citep{mateo95}, for a field away to the east of  the globular cluster M54 (which marks the nominal centre of Sgr),  showed a faint main-sequence turn-off corresponding to a `moderately old' dominant population, estimated to have age around 4~Gyr younger than that of a typical old globular cluster and a   mean metallicity of $\sim -1$~dex (plus hints of a population of blue stragglers). The then-current stellar evolution models estimated  14~Gyr for a classical halo globular cluster, yielding $\sim 10$~Gyr for the field stars of Sgr.   A more detailed analysis \citep{layden1995} of the evolved stars in two fields, one including  M54,  demonstrated that the globular cluster was indeed at the same distance as the field stars in Sgr, and further that the RGB and HB structure of Sgr was consistent with a dominant population of age $\sim 10$~Gyr and a metallicity of [Fe/H] $\sim -0.5$, plus a minor contribution from more metal-poor stars ([Fe/H] $\sim -1.3$) of similar or older age. The stellar content of the cluster M54 is complex, showing multiple populations \citep{milonem54}, but is distinct from that of the field of Sgr, being more metal-poor and older. That the stellar population of Sgr  is dominated by stars $\sim 4-5$~Gyr younger than typical old globular clusters was confirmed by subsequent analyses of wide-area imaging data \citep{marconi1998,bellazzini-survey,bellazzini-results}. As summarised by \citet{bellazzini2006},  \lq \lq more than 80\% of Sgr stars belong to a metal-rich and old-intermediate-age population". Those authors' preferred solution is either the combination of  mean age 8~Gyr (on the scale with halo globular clusters having ages $\sim 12$~Gyr) and metallicity $ {\rm Z = 0.004,\, [M/H] \sim -0.7}$~dex, or   mean age 6.5~Gyr and metallicity $ {\rm Z = 0.008,\, [M/H] \sim -0.4}$~dex. Spectroscopic measurements of the metallicity of Sgr stars close to M54 generally  yield a peak at [Fe/H] $\sim -0.4 $~dex \citep{bellazzini2008,mucc2017,Sgr-muse}, with a slight decline with distance from M54 \citep{mucc2017}. The brighter of the two tidal streams from Sgr, identified in the `Field of Streams' of the SDSS imaging data  \citep{FoS}, also contains a dominant population that is $\sim 7$~Gyr old \citep{deboer2015}, together with older, more metal-poor stars. The main-sequence turn-off mass for age $\sim 7$~Gyr and metallicity $\sim -0.4$~dex is $\simlt 1$~M$_\odot$ \citep[][PARSEC stellar evolution tracks]{parsec2017}, within the range of `solar-type' stars whose close-binary fraction was derived by \citet{Moe2018}, namely $\sim 0.6 - 1.5$~M$_\odot$. }

{ The colour-magnitude diagram derived from the deep photometric data contain a clear `blue plume' that could be either blue stragglers or genuinely young stars, with ages less than 1~Gyr. \citet{mucc2017} derive a wide range of photometric metallicities for \lq blue plume' stars, from below one-tenth of solar to above the solar value, which would be surprising for a true very young population. They further found that the radial profile of the \lq blue plume' was indistinguishable from  that of the intermediate metallicity, dominant population of Sgr, albeit that this is restricted to the region within 2.5~arcmin ($\sim 20$~pc) of the centre of M54 (their Figure~9). The wide-area photometric data of \citet{bellazzini-survey} showed that the `blue plume' stars followed the radial profile of the dominant older population (traced by HB and RGB stars) over the several square degrees of their survey fields (their Figure~14; note that 1~degree subtends $\sim 500$~pc at the distance of the core of Sgr). Again, this is not expected for very young stars but is consistent with the \lq blue plume' stars being blue stragglers  associated with this dominant older population. }

{  This  dominant older population in Sgr is $\sim 4$~Gyr younger than that of the less luminous and more metal-poor galaxies. \citet{Momany2007} measure the blue straggler population relative to the horizontal branch population. As noted by \citet{renzini1994}, the duration of the HB for ages older than a few Gyr (so that the stars experienced the helium flash at the tip of the RGB) is only weakly dependent on age. The scaling in  \citet[his equation~(4)]{renzini1994} predicts a difference in duration of the HB of $\sim 1$\% between ages 7~Gyr and 10~Gyr, which results in negligible variation in the normalisations of the BSS fraction.   The duration of the blue-straggler phase for a younger population may be expected to be shorter than that for an older population, reflecting the larger turn-off mass. This would lower the relative frequency of BSS to HB. Should the duration scale simply with the lifetime of the most-massive main-sequence star then the observed ratio for Sgr would be a factor of $10/7 = 1.4$ below that expected for a 10~Gyr old population. There is also a slight compensatory tendency for the close-binary fraction to be higher (by $\sim 15$\%) for 1~M$_\odot$  primaries than for 0.8~M$_\odot$  primaries, at fixed metallicity \citep{Moe2018}.   That the point for Sgr does lie below the trend followed by the older populations in Fig.~2 is interesting, but should not be over-interpreted, as  the Poisson uncertainties in the star counts are significantly larger than these effects. }

\subsubsection{{The Galactic Bulge}}

We have extended the metallicity range under consideration even further 
by including the Galactic bulge, for which the BSS population may be identified on the basis of location in the CMD, as for the external galaxies.

{The ability to constrain the age and
metallicity distributions of bulge stars through their CMD is limited by the reliability
of the subtraction of foreground disc stars. The identification of disc stars is most robustly achieved through their kinematics, specifically the signature of close-to-circular orbits in their proper motions. Recent analyses of fields in the bulge  utilised two epochs of imaging with the {\it Hubble Space Telescope\/} (HST), facilitating the removal of foreground stars through their significant positive proper motions in longitude \citep{clarkson2011,renzini2018,bernard2018}. }

A  field with line-of-sight passing $\sim 350$~pc from the Galactic centre was studied
with the ACS WFC {(F606W and F814W filters)} on HST by  \citet{clarkson2008,clarkson2011}. {The data were obtained in two epochs and were designed to detect photometric variability.}   The
bulge stars were separated from foreground disc stars using on the basis of
proper motions, and the dominant remaining (bulge) population in the colour-magnitude
diagram is well fit by an old (age $\sim 11$~Gyr), {solar-metallicity isochrone (see their Fig.~1). There is a scattering of stars bluer than the dominant old turnoff, occupying the locus of metal-poor ([Fe/H] $ \sim -1$), young (age $\simlt 5$~Gyr) populations \citep[see  Figure~20 of][]{clarkson2008}.  A significant fraction, $\sim 10$\%,  of these blue stars  showed light curves characteristic of W Uma contact binaries, while  no such light curves were seen for stars with disc kinematics  \citep{clarkson2011}. Those authors estimate that their time-series data are sensitive to $\sim 35$\% of the parent W UMa binary population, increasing the fraction of blue stars in such systems to $\sim 25$\%. Further correcting for the underlying binary orbital period distribution, guided by the data for solar-neighbourhood stars, led to the conclusion that all the blue stars could be BSS in binaries.\footnote{This inferred population of BSS would closely resemble that in the open cluster NGC~188, which also has $\sim 20$\% in short-period binaries \citep{mathieu2009}.} More conservatively, \citet{clarkson2011} estimate that   at most $\sim 3.5$\%  of the bulge in their field is a genuinely young population, defined as ages $\simlt 5$~Gyr. }

{\citet{renzini2018} analyzed the CMDs for the set of fields targeted for the {\it Hubble Space  Telescope\/} WFC3 Galactic Bulge Treasury program and found age distributions in excellent agreement with \cite{clarkson2011}, namely a dominant old population and $\sim 3$\% possibly as young as, or younger than, $\sim 5$~Gyr.  These fields again have two epochs of data, facilitating removal of foreground disc stars through cuts in proper motion. \citet{renzini2018} derived a photometric metallicity for each star, and investigated the relative ages of the most metal-rich ([Fe/H] $\simgt +0.2$~dex) and most metal-poor ([Fe/H] $\simlt -0.7$~dex) stars by comparing their luminosity functions; both were consistent with ages $\sim 10$Gyr. }

{This same WFC3 Galactic Bulge Treasury dataset was analysed by \citet{bernard2018} using a different approach, namely direct modelling  of synthetic (I, V-I) CMDs from linear combinations of a wide range of input Simple Stellar Populations (SSPs) of assumed age and metallicity. Their best-fit  model is dominated by the SSP with solar metallicity and age 11.5~Gyr (see bottom right panel of their Fig.~9). There are non-negligible contributions from younger SSPs, particularly those with super-solar metallicities. Indeed, in their \lq\lq cleanest'' composite stellar population,   $\sim 30$\% of the stars above solar metallicity have ages less than 8~Gyr, decreasing to $\simlt 10$\% at lower metallicities. It should be noted that blue stragglers are not included in the SSPs, and are an alternative possibility for the presence of blue stars beyond the dominant main-sequence turn-off. The quoted fractions of younger stars are thus strictly upper limits.  That said, the fractions of younger stars found by \citet{bernard2018} are consistent with the conclusions of \citet{bensby2017} who obtained and analyzed spectra of microlensed stars, likely dwarf and subgiant stars within $\sim 0.5-1$~kpc of the Galactic Center. They derived most-probable ages for individual stars by fitting to isochrones and found, after correcting for an apparent bias towards younger and more metal-rich stars, that above solar iron abundance about one-third of the stars were younger than 8~Gyr and in the metallicity bin between $-0.5$~dex and solar, one-fifth are younger than 8~Gyr. They conclude that around 15\% of the total population in the bulge could be younger than 5~Gyr. Again, BSS are not considered in their analysis. }

{There is thus a concensus that the dominant population in the bulge is older than 8~Gyr, with estimates of the fraction younger than this ranging from 0\% to $\sim 25$\%. There is clearly a well-defined main-sequence turn-off, plus `blue plume'. The analysis of \citet{clarkson2011} is the only study with  time-series photometry, which enabled them to determine the variability of the `blue plume' stars and thus to quantify their binary nature and determine that a significant fraction are indeed blue stragglers.  }

  {The  core-helium-burning phase in the (solar-metallicity) bulge population consists predominantly of Red Clump (RC) stars and the BSS fraction  is given relative to the sum of BHB, RHB and RC.  Taking account of uncertainties in the counts of HB stars, }  
 \citet{clarkson2011} report their estimated ratio of F$^{BSS}_{HB}$ to be in the range of $18/58 = 0.31$ to $37/30 = 1.23$. We have indicated the corresponding range in log(F$^{BSS}_{HB}$) in Fig.~\ref{fig:table2}, using a distinct symbol (star) for the mean value in this bulge field,  as it was not part of the uniform analysis of
\citet{Momany2007} that forms the rest of the sample plotted.

\subsubsection{Willman~1}
\label{sec:Wil1}

{We excluded Willman~1 from Table~\ref{tab:santana} due to its  uncertain dynamical state - the derived systemic velocity for the  innermost candidate member stars is significantly offset from the  outer members \citep{willman2011}. Indeed, as discussed in \citet[][their section 5.2]{willman2011}, the systemic velocity falls off by $\sim 8~$km/s from the centre to the outskirts, with a steep drop within one half-light radius, while the velocity dispersion about the systemic velocity is formally $0 \pm 2.5$km/s when calculated in three radial bins.  Further, the estimated mean  metallicity ([Fe/H] $\sim -2$~dex) is based on only three stars with a large dispersion in  metallicity.  The unknown dynamical history, and  unknown present state, of Willman~1  makes any attempt at quantification of the BSS  population very difficult to interpret. Furthermore, the very small number of (candidate) member stars in this extremely faint system ($M_V \sim -2$ \citealp{willman2011}) leads to such large uncertainties in the fraction of BSS that it would not provide a meaningful addition to the present sample (for reference, the value derived by \citealt{santana2013} is $F^{BSS}_{RGB} = 0.85 \pm 0.75$).}

\subsubsection{Carina dSph}
\label{sec:carina}

{We excluded the Carina dSph from Table~\ref{tab:momany}.  This galaxy has a very unusual colour-magnitude diagram, featuring two well separated and  well populated subgiant branches, and a third less distinct subgiant branch \citep[e.g.][]{tammy-carina,carina}. The derived star-formation history is episodic, with two, approximately equal mass, bursts occuring firstly more than 10~Gyr ago and then around 7~Gyr ago,  with a weaker episode some 2-3Gyr ago \citep[e.g.][including earlier results in their Table~4]{jen-carina1}. These existence of these multiple populations, each with a different characteristic metallicity, and the lack of one clearly dominant one, make it essentially impossible to associate the detected BSS (which is the possible fourth population in \citealt{jen-carina1}) with the  parent population.}

\begin{table*}
\caption{Relative population of blue stragglers to red-giant branch stars, for the sample of  \citet{santana2013}, together with mean stellar iron-peak metallicities (column 5) and absolute V-band magnitudes (column 7), with associated errors, taken from the compilation of  \citet{AMcC2012}. Galaxies are ordered by (increasing) mean metallicity.  The values in columns (1) and (2) are as reported by \citet{santana2013}, while those in columns (3) and (4) were recalculated as described in the text.}
\label{tab:santana}
\centering 
\begin{tabular}{l c c c c c c c c}
  \hline
\rule{0pt}{1ex} & & & & & & &  & \\
Galaxy & F$^{BSS}_{RGB}$  & error  & log(F$^{BSS}_{RGB}$)  & error  & <[Fe/H]> & error & $M_V$ & error \\
& (1) & (2) & (3) & (4) & (5) & (6) & (7) & (8)  \\

\rule{0pt}{1ex} & & & & & & \\
\hline
\rule{0pt}{0.1pt} & & & & & & \\
 Coma Berenices & 0.49 & 0.33 &  $-0.31$ & 0.31 & $-2.60$ & 0.05 & $-4.1$ & 0.5 \\
 Bo\"otes I  & 0.36 & 0.07 & $-0.44$ & 0.08 & $-2.55$ & 0.11 & $-6.3$ & 0.2 \\
 Ursa Major II  & 0.31 & 0.09 & $-0.51$ & 0.13 & $-2.47$ & 0.06 & $-4.2$ & 0.6 \\
 Hercules  & 0.26 & 0.07 & $-0.59$ & 0.12 & $-2.41$ & 0.04 & $-6.6$ & 0.4 \\
 Ursa Major I  & 0.28 & 0.12 & $-0.55$ & 0.19 & $-2.18$ & 0.04 & $-5.5$ & 0.3 \\
 Ursa Minor & 0.3 & 0.02 & $-0.53$ & 0.02 & $-2.13$ & 0.01 & $-8.8$ & 0.5 \\
 Segue II  & 0.26 & 0.22 & $-0.62$ & 0.39 & $-2.00$ & 0.25 & $-2.5$ & 0.3 \\
 Canis Venatici I  & 0.31 & 0.03 & $ -0.50$ & 0.04  & $-1.98$ & 0.01 & $-8.6$ & 0.2 \\
 Sextans  & 0.29 & 0.02 & $-0.54$ & 0.02 & $-1.93$ & 0.01 & $-9.3$ & 0.5 \\
 Draco  & 0.27 & 0.02 & $-0.58$ & 0.02 & $-1.93$ & 0.01 & $-8.8$ & 0.3 \\
  Bo\"otes II  & 0.26 &  0.17 & $-0.69$ & 0.39  & $-1.79$ & 0.05 & $-2.7 $ & 0.9 \\
\rule{0pt}{0.1pt} & & & & & & \\
\hline
\end{tabular}
\\
\end{table*}

\begin{table*}
\caption{Logarithm of the relative frequency of blue stragglers to horizontal-branch stars (being the sum of RHB and BHB stars) for the sample of \citet{Momany2007}, tabulated in \citet{Momany2015}. The  mean stellar iron-peak metallicities (column 3) and absolute V-band magnitudes (column 5) and associated errors are taken from the compilation of  \citet{AMcC2012} and galaxies are ordered by (increasing) mean metallicity.}
\label{tab:momany}
\centering 
\begin{tabular}{l c c c c c c}
  \hline
\rule{0pt}{0.1pt} & & & & & & \\
  Galaxy & log(F$^{BSS}_{HB}$) & error  & <[Fe/H]> & error & $M_V$ & error \\
  & (1) & (2) & (3) & (4) & (5) & (6)   \\
  \rule{0pt}{0.1pt} & & & & & & \\
\hline
\rule{0pt}{0.1pt} & & & & & & \\
 Bo\"otes I  & 0.26 & 0.15 & $-2.55$ & 0.11 & $-6.3$ & 0.2 \\
 Ursa Major I  & 0.20 & 0.18 & $-2.18$ & 0.04 & $-5.5$ & 0.3 \\
 Ursa Minor & 0.13 & 0.13 & $-2.13$ & 0.01 & $-8.8$ & 0.5 \\
 Sextans  & 0.05 & 0.12 & $-1.93$ & 0.01 & $-9.3$ & 0.5 \\
 Draco  & 0.09 & 0.15 & $-1.93$ & 0.01 & $-8.8$ & 0.3 \\
 Leo II  & $-0.01$ & 0.09 & $-1.62$ & 0.01 & $-9.8 $ & 0.3 \\
 Sculptor & 0.01 & 0.13 & $-1.68$ & 0.01 & $-11.1 $ & 0.5 \\
  Sagittarius & $-0.26$ & 0.26 & $-0.4$ & 0.2 & $-13.5 $ & 0.3 \\
  \rule{0pt}{0.1pt} & & & & & & \\
  \hline
\end{tabular}
\\
\end{table*}

\subsection{The Dependence on Mean Metallicity}
\label{subsec:feh}

It is now well-established that there is a tight correlation between
the mean metallicity of a dSph galaxy and its luminosity
\citep{Norris2010,Kirby2013}. This is most likely a causal relation, reflecting the systematic
effects of stellar feedback and gas loss during star formation \citep[e.g.][]{dekel1986}. This correlation implies that the (mild) anti-correlation between the host galaxy's luminosity and BSS fraction found by \citet{Momany2007} actually reflects an underlying trend between metallicity and BSS fraction.  Thus,  given the 
finding by \citet{Moe2018} that there is a (plausibly causal)
anti-correlation between the close-binary fraction for low-mass stars and
mean metallicity of the parent population,   there should be an anti-correlation between  metallicity and  BSS
fraction in dSph galaxies (and indeed in any system where the primary mode of BSS formation is mass transfer in close binaries). 

\begin{figure}
  	\includegraphics[width=0.75\columnwidth,angle=270]{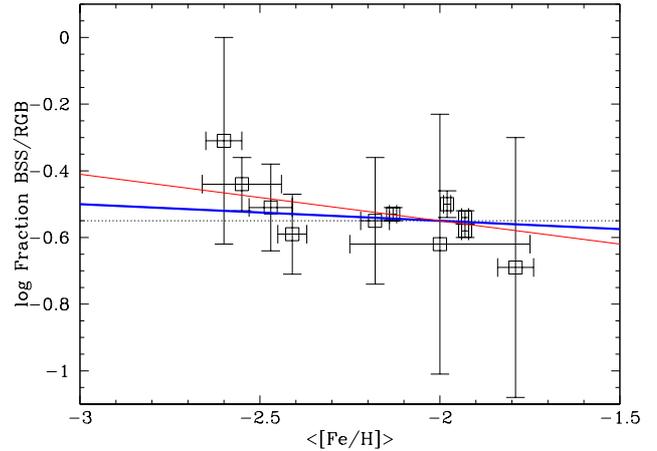}
    \caption{The fraction BSS/RGB (plotted as a logarithm) for the sample in Table~\ref{tab:santana} (points), plotted as a function of the mean stellar iron-peak abundance of the host galaxy. The uncertainties in log(F$^{BSS/RGB}$) have been calculated as described in the text. The dotted horizontal line indicates the constant value of $-0.55$ that is consistent with the data. The thin (red) line  is the best-fit non-constant trend and provides a lower value of  reduced $\chi ^2$ than a constant value. This shows that a decline in BSS fraction with increasing metallicity is favoured by the data.  The thick (blue) line follows the slow decrease found  by \citet{Moe2018} to fit the trend for close-binary fraction over this metallicity regime, arbitrarily normalised. This  has a slightly shallower slope than the best-fit decline for the BSS fraction.}
    \label{fig:table1}
\end{figure}

Fig.~\ref{fig:table1} shows the data for the (log of the) BSS fraction
relative to RGB stars from \citet{santana2013}, plotted against the
mean iron-peak abundance of the host galaxy (taken from the compilation of
\citealt{AMcC2012}). A constant value of log(F$_{RGB}^{BSS}$),
independent of metallicity, is consistent with the data, providing a
reduced $\chi^2/\nu = 7.6/10=0.76$ (the dotted line in
Fig.~\ref{fig:table1}). Allowing a non-zero slope provides a lower
value of the reduced  $\chi^2$, with the best-fit slope of $\Delta$log(F$_{RGB}^{BSS}$)/$\Delta$[Fe/H]$ = -0.14 \pm 0.08$ per dex of metallicity (thick red line in Fig.~\ref{fig:table1}) providing a
reduced $\chi^2/\nu = 4.9/9=0.54$. A slow decline in relative BSS population with increasing metallicity is therefore (weakly) favoured by the data.  A slight anti-correlation with metallicity, such as found by \citet{Moe2018} for close-binary fraction over this metallicity range (declining by a factor of 1.3
between $-3$~dex and $-1$~dex), and indicated (with arbitrary normalisation) by the thin blue line in Fig.~\ref{fig:table1},  thus provides a better match to the BSS data than does a constant value.

\begin{figure}
	\includegraphics[width=0.75\columnwidth,angle=270]{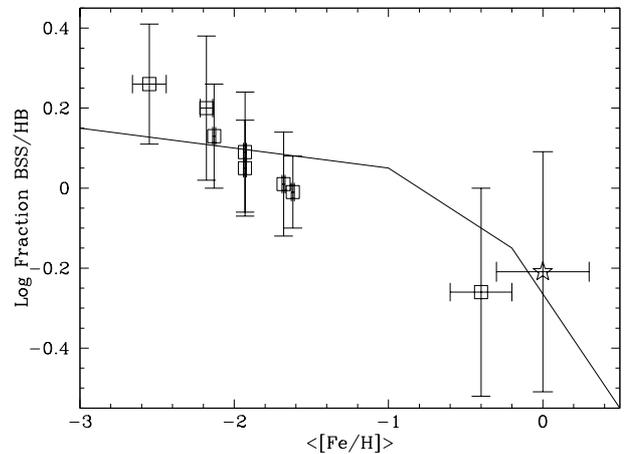}
    \caption{The logarithm of the fraction BSS/HB for the sample in
      Table~\ref{tab:momany} (open square symbols) against mean stellar iron-peak abundance, plotted with the
      relation for close binary fraction from \citet{Moe2018},
      arbitrarily normalised (line). The star symbol corresponds to
      the estimated fraction for the Galactic bulge field studied with
      the {\it Hubble Space Telescope}\/ by \citet{clarkson2011}.  The
      normalisation for the BSS fraction in this bulge field, where the main
      population has close to solar iron abundance, includes also the
      Red Clump, thus encompassing the entire core-helium burning phase (thus HB =  BHB plus RHB plus
      RC).}
    \label{fig:table2}
\end{figure}

Fig.~\ref{fig:table2} shows the data for BSS fraction relative to HB
stars (log scale) from \citet{Momany2007}, again plotted against the mean
stellar iron-peak abundance  of the host galaxy. {As discussed above, the} dSph sample in this case includes the Sagittarius dSph, the most metal-rich such system.  As in Fig.~\ref{fig:table1}, the line follows the relation
between close-binary fraction and metallicity found by
\citet{Moe2018}, once again arbitrarily normalised and plotted as straight-line segments.  

{The trends seen in both Figures~1 and 2 for the BSS fraction in the satellite galaxies (plus the Galactic bulge) as a function of metallicity are consistent with the trend found by \citet{Moe2018} for the close-binary fraction for solar-type stars in the field of the Galaxy.} 

\section{{From Close Binaries to BSS}}

A direct comparison between the trend seen for the BSS data and that
for close binaries is valid provided mass transfer in close binaries
is the dominant means by which BSS form, and there is no additional
dependence on metallicity in the process. There are two simulations of
BSS formation from close binaries reported in the literature that use
the same underlying stellar evolution code, developed from the
original code of \citet{ppe1971}, adapted for binary evolution, in
different metallicity regimes: metal-poor, $Z = 0.0003$
\citep{jiang2017} and metal-rich, $Z=0.02$ \citep{chen2009}. In both
these analyses, the values of the parameters of the Monte Carlo
simulations that form the basis of the binary population synthesis are
assumed to equal those of field stars, and the internal dynamics of
star clusters are not modelled. We took the reported results from the
simulations described by Set~1 in Table~1 of \citet{chen2009} and by
Set~1 in Table~2 of \citet{jiang2017}, which use the same assumptions
about initial binary mass-ratios and separations, and estimated the
ratio of BSS (blue-sequence only, as would be identified in the
observed CMDs) to initial binary systems, at an age of $~13$~Gyr. We
find a frequency of BSS, expressed as the logarithm of the number of
BSS produced by an initial population of $10^5$~binaries, of
${\rm  log(}N_{BSS}{\rm )} = 1.49 \pm 0.06$ for the metal-poor case, and
${\rm log(}N_{BSS}{\rm )} = 1.36 \pm 0.10$ for the metal-rich
case. There is therefore a slightly higher efficiency of production for BSS
in more metal-poor populations (albeit consistent with no change, given Poisson uncertainties), formally an increase by a factor of
1.35 over a decrease of a factor of 67 in metallicity from solar to
[Fe/H] $\sim -1.7$. Thus the metallicity dependence in the
transformation of close binaries into BSS is significantly weaker than
the factor of 2.3 observed for the decrease in close binary frequency
over the same metallicity range \citep{Moe2018}. The sense of this additional
metallicity sensitivity is such that the close-binary trend with
metallicity should underestimate the observed trend of BSS with
metallicity (as is indeed the case in Fig.~\ref{fig:table1} and Fig.~\ref{fig:table2}).

{The duration of the HB phase of low-mass stars increases as  metallicity (and helium content) increases. Conversely, the duration of the RGB phase decreases with increasing metallicity (and helium content). \citet{renzini1994} quantifies these effects in his equations (1) and (2), respectively.  These scalings, for $\Delta Y/\Delta Z \sim 2$, formally predict that the RGB duration decreases by a factor of 8\% over the 0.8~dex in metallicity range of the data in Figure~1. Assuming the RGB counts are proportional to the duration of that phase, this would produce a (slight) flattening of the observed trend of the data compared to the underlying trend of the frequency of BSS. The  galaxy data would remain compatible with the data for close binaries. The wider metallicity range of the sample in Figure~2 leads to a prediction that the HB duration increases by $\sim 40$\% across the sample.  This would produce a steepening away from the underlying trend of BSS, which could provide an explanation for the steeper trend  of the galaxy data in Figure~2, compared to the relationship followed by close binaries. Even after accounting for these variations in the HB duration, metal-poor galaxies with [Fe/H] = -2.0 have a factor of two higher  occurrence rate of BSS than have the metal-rich Sgr dwarf  and Galactic bulge, consistent with the trend for close binaries.}

As noted in Section~\ref{sec:Introduction}, the most recent analysis of the BSS fraction in the field stellar halo and thick disc of the Milky Way \citep{santucci2015} was based on the SDSS spectroscopic data for A-type stars. Those authors  derived estimates of the specific frequency of BSS to {\it blue\/} HB stars only.  The lack of RHB in the normalisation (unavoidable, given the selection of A-type stars) will be expected to result in an overestimation of the fraction of BSS, particularly for the thick disc, given its higher mean metallicity  - see for example the red HB morphology of the metal-rich globular cluster 47~Tuc \citep[e.g.][]{47tuc}. The very different technique used in the derivation of the relative population of BSS adopted in this analysis argues against including its results in Fig.~\ref{fig:table2}, even as an upper limit.

Recently \citet{spencer2018} used multi-epoch radial velocity measurements to derive estimates of the binary fraction of stars on the RGB in seven of the dSph analysed in this paper. They found no trend with metallicity (their Fig.~14). However, there is only a weak trend expected over the range of metallicity covered by the sample (below $-1$~dex). Further, the inferred distribution of separations of the binaries (their Fig.~5, panel (B)) extends well into the wide-binary regime where a weaker trend is both predicted \citep{Moe2018} and observed in the Milky Way \citep{elbadry2019}.

\section{Old Open Clusters}

As noted in Section~\ref{sec:Introduction}, the BSS in old open
clusters appear to have formed {\it both\/} via mass-transfer in primordial
binaries, plausibly with orbits modified through dynamical
interactions, and via dynamically induced mergers. {It may therefore be expected that interpreting the BSS population in terms of the initial binary population would be far from straightforward. Indeed, even determining an estimate of the relative BSS fraction requires CMDs that have had the foreground/background field stars of the disc removed (most clusters are at low Galactic latitudes), ideally through a kinematic separation. The astrometric data from Gaia should alleviate this situation,  by providing the  reliable membership probabilities that are critical to a robust analysis. }

{The old open cluster Berkeley~17 illustrates the limitations of analyses prior to the release of {\it Gaia\/} DR2 \citep{GDR2}. This cluster was included in the   catalogue of \citet{AL2007}, who identified  BSS candidates for a sample of 427 open clusters. These authors found a total of 31  BSS in Berkeley 17, but subsequently all but 2 of these candidates were rejected, as non-members, by \citet{Berkeley17}, who  used both proper motion and parallax information from {\it Gaia\/} DR2 to remove field contamination. \citet{Berkeley17} did identify a significant number of BSS (23) that met the astrometric criteria for membership (21 of which were not identified by \citep{AL2007}, albeit using a different filter set to define the CMD). \citet{Berkeley17} compared the (projected) radial profile of this new sample of BSS to that of the (member) RGB stars (selected down to the same magnitude limit as the BSS), finding  the  BSS to be more centrally concentrated  - as expected for more massive stars if the cluster is dynamically evolved. 

}

Further insight can be gained from the {relatively nearby ($ d \sim 1.8$~kpc)} and populous old open cluster NGC188, which  has been the focus of significant
theoretical and observational effort into identifying the BSS
population and determining their dominant production
mechanism.  {The spectroscopic survey of \citet{geller2008} allowed removal of field disc stars through both proper motions \citep{platais2003} and line-of-sight velocities.} The cleaned CMD contains 21 BSS (using the \citealt{geller2008} colour-magnitude cuts) and  66 RGB, giving a ratio of  F$^{BSS}_{RGB} = 0.32$.  We attempted to reproduce more closely the BSS and RGB selection criteria of \citet{santana2013}, transforming the photometric data to SDSS $g,r$ following \citet{sdss2006}, and obtained 
F$^{BSS}_{RGB} = 0.27$. NGC~188 has metallicity just below the solar value, so extrapolating from the dSph values in Fig.~\ref{fig:table1} using the dependence of close-binary fraction on metallicity in \citet{Moe2018} would predict a ratio  F$^{BSS}_{RGB} \simeq 0.12$.  This factor of $0.27/0.12 \simgt 2$  discrepancy could reflect an enhanced formation rate of BSS, plausibly due to the cluster environment. {Indeed, \citet{cohen2020} found that the binary fraction of solar-type stars on the main sequence in NGC~188 is significantly higher (by a factor of $\sim 1.6$) than in the local field.} The combined effects of mass segregation (BSS would preferentially sink to the inner regions of clusters) plus Galactic tides (acting to remove stars from the outer regions of clusters) will also increase the measured fraction of BSS. 

\citet{gosnell2015} conclude, on the basis of the 
frequency of 
detection  of WD companions to BSS
in NGC188, that a lower limit of one-thirds of the BSS could have formed by mass transfer in (primordial?) binaries. Counting the additional BSS in binaries leads to an estimated 2/3 formed by mass transfer.  {\citet{geller2011} conclude, from analysis of the (narrow) binary mass-ratio, that the long-period blue straggler binaries likely formed through mass transfer processes.}  Extant detailed N-body simulations \citep[e.g.][]{geller2013} under-predict the BSS population, but this is dependent on the orbital properties assumed for primordial binaries and on the detailed treatment of the  mass transfer.

The data from {\it Gaia\/} should lead to  more complete and unbiased samples of BSS in Galactic open clusters and allow a robust investigation into possible trends with metallicity,  albeit that internal dynamical processing, mass segregation of the heavier BSS and Galactic tides will all modify the frequency of BSS in open clusters compared to  those in low-density, field environments.

\section{Discussion}
\label{sec:discussion}

The data in each of  Fig.~\ref{fig:table1} and Fig.~\ref{fig:table2} are consistent with 
a decline in the specific frequency of BSS in dSph satellite galaxies of the Milky Way, and in the Galactic bulge, with increasing metallicity
of the parent population. In all cases the fractions were derived based on star counts within a CMD. The two datasets differ in their measure of
the underlying old population: that in Fig.~\ref{fig:table1} adopts
the RGB while that in Fig.~\ref{fig:table2} adopts the HB. All of the
data pertain to field stars, rather than cluster members. 

The relatively low
stellar densities of these galaxies (especially compared to globular
clusters) favour a channel of BSS formation that derives from mass-transfer in a
primordial population of close binaries
\citep[e.g.][]{Mapelli2009}. Provided that the properties of the primordial binaries, and the rate of evolution into
BSS, are uniform within each of the two samples, the fraction of BSS
should reflect (likely proportional to) the fraction of (parent) close binary systems. Further,
the underlying populations in all cases are old, so that it is the
close binary fraction of low-mass stars that is relevant.  Thus the
strong similarity between the trends with metallicity of the BSS 
fraction in external galaxies (and in the bulge) on the one hand, and of the  close-binary fraction for low-mass stars in the nearby field stars in the Milky Way - the mean trend of which is represented by the solid lines in Fig.~\ref{fig:table1} and Fig.~\ref{fig:table2} - on the other, suggests the same underlying physical mechanism.

{As noted earlier, the relative normalisations between the trends for close binaries and the data points for the galaxies in Figures~1 and 2 were chosen arbitrarily. It is not at all straightforward to extract from this what fraction of close binaries become blue stragglers. The reported fraction of blue stragglers is measured relative to an evolved phases of single stars (RGB or HB) of a narrow range of ages with a narrow mass range, while the fraction of close binaries (specified in terms of separation), with primary star within a specified mass range ($\sim 0.6-1.5$~M$_\odot$), is measured relative to all single stars in that mass range \citep{Moe2018}. There is an additional factor  due to the fact that the lowest-mass blue stragglers within any one galaxy are likely fainter than the dominant MSTO (note that for an old population, ages $\simgt 10$~Gyr, the MSTO mass varies by only $\sim 0.15$~M$_\odot$ over the range of metallicity shown in Figures~1 and 2;  \citealp{parsec2017}). Quantification of this factor depends on the detailed stellar initial mass function. We choose to simply absorb it into the arbitrary normalisation, making the implicit assumption of an invariant IMF. As discussed in \citet{Moe2018},  the increased probability that they find  for metal-poor discs to fragment must alter the IMF. However, they estimate the net effect to be that extremely metal-poor binary systems should,   on average, be $\sim 30$\% more massive than their metal-rich counterparts. This difference in expected characteristic (binary system) mass is within the uncertainties of derived mass functions \citep[][and references therein]{kroupa}. }

 The model of \citet{Moe2018} predicts a metallicity dependence for the binary fraction of low-mass stars only for those small scales on which the fragmentation is driven by gravitational instability of individual proto-stellar discs, rather than the larger-scale fragmentation of turbulent cores. {The trend for higher close-binary fraction at lower metallicities held regardless of whether the metal-poor population was in the disc or in the halo.} Interpreting the trends found in this paper for the BSS fraction in  external galaxies in terms of close-binary fraction, the similarity between the results for for field stars in the dSph and field stars in our Galaxy supports the proposal that the small-scale star-formation process has universal physical underpinnings, independent of larger-scale environment.
 
\section*{Acknowledgements}

{We thank the referee for their constructive comments.} RFGW thanks her sister, Katherine Barber, for support and Imants Plaitis for helpful  discussions. RFGW and KK thank the Research School of Astronomy and Astrophysics of the Australia National University for hospitality during their visits in August 2018, when this collaboration was conceived. RFGW's visit to ANU was supported by ARC Discovery Project grant DP160103737.  RFGW thanks the Kavli Institute for Theoretical Physics (KITP)  and the Simons Foundation for  support as a Simons Distinguished Visiting Scholar.  KITP and this research were supported in part by the National Science Foundation under Grant No. NSF PHY-1748958. MM and KK acknowledge financial support from NASA under Grant No. ATP-170070. 




\appendix

\section{Galaxies with both measures of the specific frequency of BSS}

\begin{figure}
	\includegraphics[width=0.75\columnwidth,angle=270]{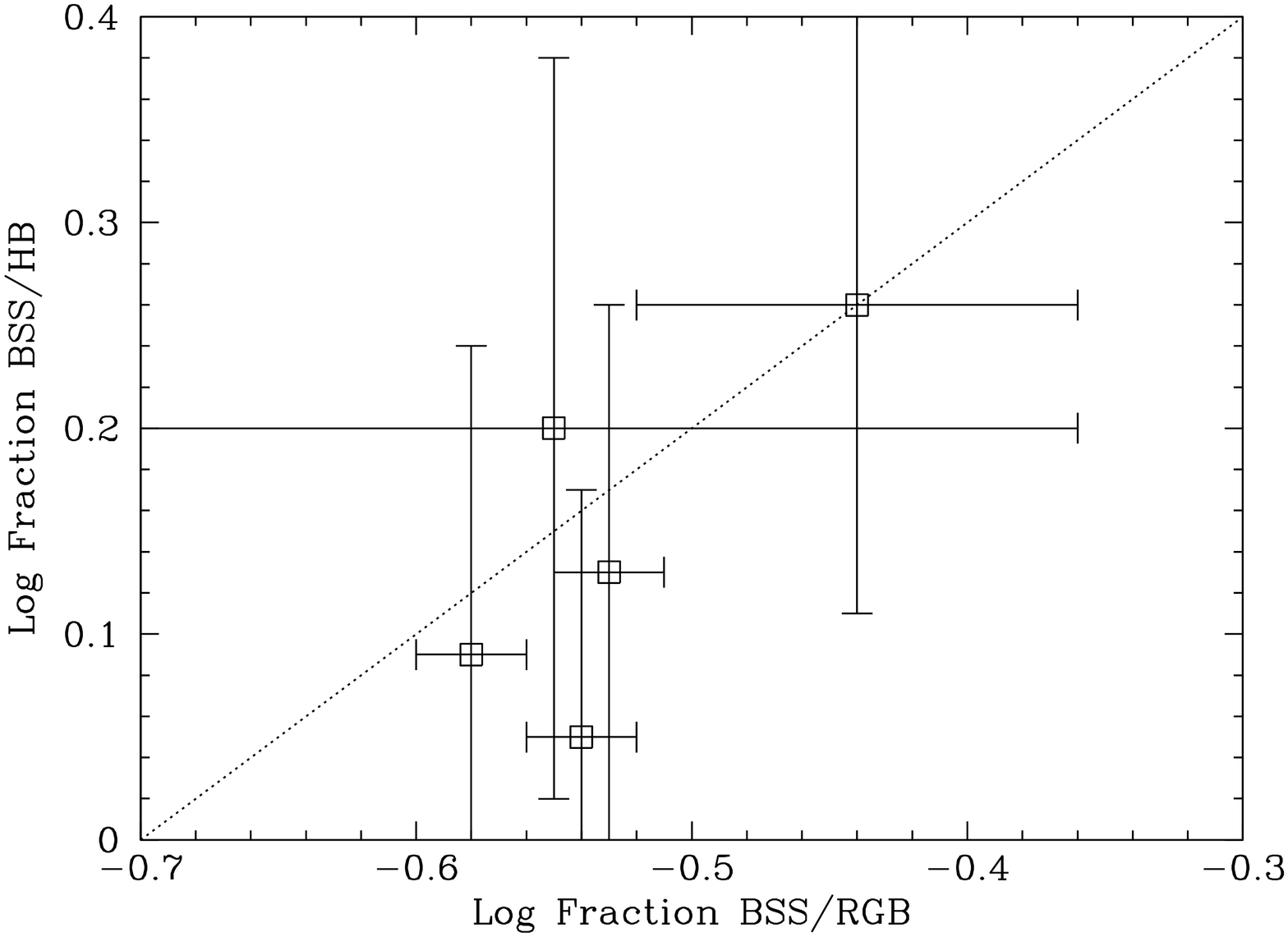}
  \caption{The two measures of the specific frequency of BSS for the five galaxies with both. The dotted line indicates a one-to-one relation. While the uncertainties are large, the points do scatter about the one-to-one line. }
      \label{fig:both}
\end{figure}

There are five galaxies in common between the samples of
\citet{Momany2007} and \citet{santana2013}, covering a (moderate)
range in mean metallicity from $-1.93$~dex to $-2.55$~dex. The two
measures of the specific frequency of blue stragglers  for each galaxy are clearly not equal, as can be read from
the entries in Tables~1 and 2 above. They should, however, be strongly
correlated as they purport to measure the same underlying quantity.
The two estimates for the relative BSS population for  these five
galaxies are plotted against each other in Fig.~\ref{fig:both}.  While the uncertainties are large, the points do scatter about the one-to-one line. {The most discrepant point is that representing the Sextans dSph; as may be seen in Tables~1 and 2, Sextans and Draco are very similar in terms of BSS relative to RGB, but differ in terms of BSS relative to HB, with Sextans in this case having a lower reported frequency of BSS. The large uncertainties render this nominal offset barely significant.} 

\bsp	
\label{lastpage}
\end{document}